# Global N-body Simulation of Gap Edge Structures Created by Perturbations from a Small Satellite Embedded in Saturn's Rings II: The Effect of Satellite's Orbital Eccentricity and Inclination

Naoya Torii[a,b], Shigeru Ida[b], Eiichiro Kokubo[c] and Shugo Michikoshi[d]

[b]*Department of Earth and Planetary Sciences, Institute of Science Tokyo, Ookayama, Meguro-ku, Tokyo, 152-8551, Japan*
[b]*Earth-Life Science Institute, Institute of Science Tokyo, Ookayama, Meguro-ku, Tokyo, 152-8550, Japan*
[c]*National Astronomical Observatory of Japan, Osawa, Mitaka, Tokyo, 181-8588, Japan*
[d]*Department of the Study of Contemporary Society, Kyoto Women's University, Imakumano, Higashiyama, Kyoto, 605-8501, Japan*

ABSTRACT

Small satellites, Pan and Daphnis, are embedded in Saturn's rings and opening a clear gap with satellite wakes at the gap edges. Furthermore, in the case of Daphnis, pronounced vertical wall structures casting shadows on the rings are also observed in the satellite wakes. In our previous paper (Torii, N., Ida, S., Kokubo, E., Michikoshi, S. [2024]. Icarus, 425, 116029), we found through global 3D N-body simulation that the ring particles' lateral epicycle motions excited by an encounter with a satellite are converted to the vertical motions through oblique physical collisions between the particles at the wavefronts of satellite wakes. In order to highlight this dynamics, Torii et al. (2024) considered the circular ($e_s = 0$) and coplanar ($i_s = 0$) satellite orbit, where $e_s$ and $i_s$ are the satellite orbital eccentricity and inclination, respectively. However, Daphnis has non-negligible eccentricity and inclination. In this paper, we perform a global 3D N-body simulation with non-zero $e_s$ or non-zero $i_s$ of the satellite orbit to investigate how they affect the gap edge structures. We found that the effect of satellite eccentricity is important both in the satellite wakes and the vertical walls at the gap edges. The non-sinusoidal sawtooth-like satellite wakes and azimuthally more localized vertical walls observed by Cassini are simultaneously reproduced in the detailed structures and spatial scales. Both of them periodically vary due to the satellite excursions between the apocenter and the pericenter. The ring particles in outer (inner) rings that undergo closest encounters with the satellite near the apocenter (pericenter) are excited the most highly. Because the excited eccentricities of the ring particles are converted to the inclinations through physical collisions, the conversion is the most active for the particles that acquire the highest eccentricities, resulting in the azimuthally more localized vertical wall structures. The predicted height of the tallest vertical walls is $\sim 0.2$ times the satellite Hill radius in the case of the satellite eccentricity comparable to Daphnis when adopting Hill scaling, which is twice as much as the height obtained in the case of the circular satellite orbit and is quantitatively more consistent with the Cassini observation. The simulation with the inclined satellite orbit reveals that the local vertical walls created by the particle-particle collisions persist at the satellite wavefronts and are superposed with the global bending waves induced by the satellite out-of-plane perturbations. These results show that the observed vertical walls are actually formed by the satellite wakes followed by their conversion to the vertical motions through inter-particle collisions, rather than by the out-of-plane perturbation from the satellite in an inclined orbit.

## 1. Introduction

The Voyager and Cassini observations have brought us a large number of unexpected discoveries and revealed that many spectacular phenomena are occurring in Saturn's rings. One of the discoveries is the gap structure created by a small embedded satellite, Daphnis or Pan (see Table 1 for a summary of their orbital parameters), including the satellite wakes (e.g., Porco et al., 2005; Tiscareno et al., 2019) and sharply truncated edge (Showalter, 1991). The gap structure created by Daphnis further has vertical wall structures (Weiss et al., 2009, see also Fig. 1).

The formation mechanism of these features at the gap edge has been chiefly investigated separately by different theoretical approaches such as the streamline model (e.g., Borderies et al., 1982, 1983, 1985, 1989), 1D radial diffusion

torii.n.aa@m.titech.ac.jp (N. Torii); ida@elsi.jp (S. Ida); kokubo.eiichiro@nao.ac.jp (E. Kokubo); michikos@kyoto-wu.ac.jp (S. Michikoshi)
ORCID(s): 0009-0003-5452-7473 (N. Torii)

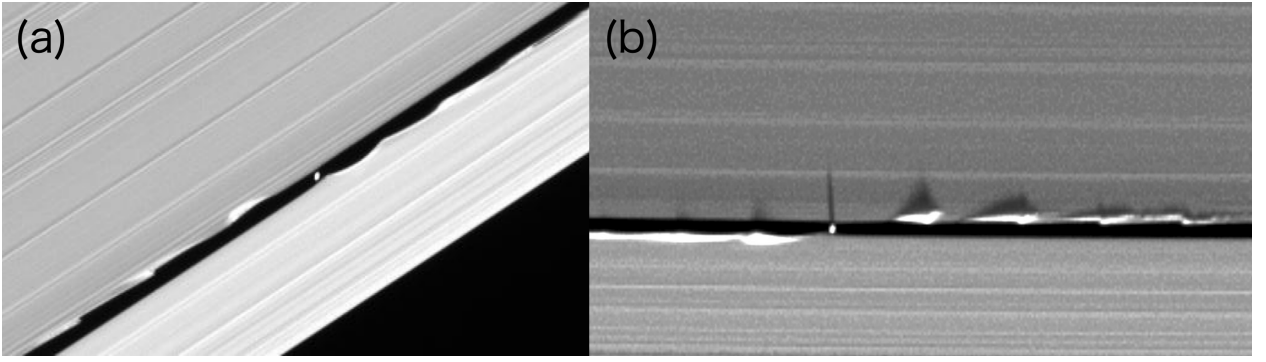

**Figure 1:** (a)Non-sinusoidal satellite wakes at Keeler gap edge. (b)The vertical structure of the satellite wakes at the Keeler gap edge with clear shadows cast on the rings. (CREDIT: NASA/JPL/Space Science Institute)

**Table 1**
Dynamical parameters of the embedded satellites, Daphnis and Pan, and the Encke and Keeler gaps (data from Weiss et al., 2009; Jacobson et al., 2008): $a_s$, $M_s/M_p$, $r_{H,s}$, $\Delta a$, $e_s$ and $i_s$ are the semimajor axis, the mass ratio to Saturn, the Hill radius of the satellite, the half-width of the gap, satellite's eccentricity and satellite's inclination, respectively, where $M_p$ is the mass of Saturn. The reference data for $e_s$ and $i_s$ in the table are from Jacobson et al. (2008) that used the Cassini imaging observations through 2007 (*We note that the observations through 2017 show a possibility that $e_s$ and $i_s$ vary with time (Jacobson (2024); V. Lainey, 2023, private communication).

| Gap | Satellite | $a_s$ [km] | $M_s$ [$M_p$] | $r_{H,s}$ [km] | $\Delta a$ [km] | $e_s$ | $i_s$ [rad] |
|---|---|---|---|---|---|---|---|
| Encke | Pan | 133,584 | $7.7 \times 10^{-12}$ | 18.3 | 161 | $(1.44 \pm 0.54) \times 10^{-5}$ | $(1.74 \pm 6.98) \times 10^{-6}$ |
| Keeler | Daphnis | 136,506 | $(1.0 - 1.6) \times 10^{-13}$ | $4.39 - 5.14$ | $13 - 20$ (inner) $14 - 16$ (outer) | $(3.31 \pm 0.62) \times 10^{-5}$ * | $(6.28 \pm 2.27) \times 10^{-5}$ * |

model (e.g., Grätz et al., 2018, 2019), local N-body simulation (e.g., Lewis and Stewart, 2000, 2005, 2009; Lewis et al., 2011; Robbins et al., 2010) and test particle model without any particle-particle interactions (e.g., Weiss et al., 2009).

Recently, we investigated these features with a global 3D N-body simulation with $N = 10^6 - 10^7$ self-gravitating ring particles and an embedded satellite in a non-inclined circular orbit and succeeded in reproducing several aspects at the gap edge simultaneously in a single simulation (Torii et al., 2024, hereafter Paper I). In particular, we proposed a new formation mechanism for vertical wall structures. Daphnis, with averaged physical radius $\sim 4\,\text{km}$, is embedded in the Keeler gap with a half-width $\sim 15\,\text{km}$ in the outermost region in the A ring. The Cassini observations found the mountain-like vertical walls of ring particles in the inner and outer gap edges. The walls are as high as $\sim 1\,\text{km}$, while the typical thickness of the A ring is only $\sim 10\,\text{m}$.

The results in Paper I are briefly summarized as follows. A satellite embedded in rings opens up a gap by its gravitational perturbations, exerting satellite wakes near the gap edge. The gap half width is predicted as $\Delta a \sim 3.5\,r_{H,s}$ where $r_{H,s}$ is the Hill radius of the satellite defined by $r_{H,s} = (M_s/3\,M_p)^{1/3} a_s$, where $M_s$ and $M_p$ are the satellite's and host planet's masses, respectively, and $a_s$ is the satellite's semimajor axis. Substituting Daphnis-Saturn mean mass ratio $M_s/M_p \simeq 1.2 \times 10^{-13}$, Daphnis' Hill radius is $r_{H,s} \simeq 3.42 \times 10^{-5} a_s \simeq 4.66\,\text{km}$ (see also Weiss et al., 2009). The predicted gap half width is $\Delta a \sim 16\,\text{km}$, which is consistent with the observation. The radial amplitude of the satellite wakes is $\sim 0.5\,r_{H,s}$ at the gap edge. Near the wavefronts of satellite wakes, inelastic collisions between ring particles damp their eccentricities excited by the satellite's perturbations, enhancing the surface density and making the gap edge sharp, which is known as negative diffusion (Lewis et al., 2011; Sickafoose and Lewis, 2024). At the same time, the collisions convert the lateral motions of ring particles to vertical ones by just a single or a few collision(s), because most of them are vertically oblique collisions. This conversion immediately forms the vertical wall structures, even if the satellite orbit is coplanar with the ring midplane and the ring thickness is as small as typical ring particle sizes. Because the initially coherent epicyclic motions deviate more in the second epicycle wake than in the first, the collisional conversion from lateral to vertical motions is more efficient in the second wake. As a result, the satellite wakes are associated with the vertical walls, and the wall is the highest ($\sim 0.1\, r_{\text{H,s}}$) in the second wakes. We note that Hoffmann et al. (2015) proposed the conversion from lateral motions to vertical motions to explain the vertical structure around embedded propeller moonlets. They argued that the conversion is a "thermal" diffusion process due to many inter-particle low-velocity collisions, which leads to a locally isotropic velocity dispersion of particles. On the other hand, we find that for a gap-opening satellite, the rapid conversion occurs through only a single or a few oblique high-velocity collision(s) right after the scattering by the satellite that excites the lateral motions of particles. This initial relaxation process is the mechanism for forming the vertical walls. Because the amplitude of the excited epicycle motions is $\sim O(r_{\text{H,s}})$, our finding naturally leads to the conclusion that the wall height is proportional to $r_{\text{H,s}}$.

In Paper I, we performed 11 simulations with $M_s/M_p = 8\times10^{-7}$ to $2\times10^{-5}$ and found that the $M_s/M_p$ dependence of the results is scaled well with $r_{\text{H,s}}$. Although the actual mass of Daphnis is $M_s/M_p = 1.2 \times 10^{-13}$, the much larger masses were adopted in the simulations for computational convenience. However, we can discuss the Daphnis case by extrapolation with this scaling: the highest part of the vertical walls is predicted as $\sim 0.1\, r_{\text{H,s}} \sim 0.47\,\text{km}$ (Table 1), which is consistent with the Cassini observation within a factor of 2.[1]

We only considered the case where the satellite orbit is circular ($e = 0$) and coplanar to the ring midplane ($i = 0$) in Paper I. However, the Cassini observation data suggest that Daphnis may have relatively large orbital eccentricity and inclination (Jacobson et al., 2008): $a_s e_s \simeq 4.5\,\text{km} \simeq 0.91\, r_{\text{H,s}}$ and $a_s i_s \simeq 8.6\,\text{km} \simeq 1.8\, r_{\text{H,s}}$. The half width of Keeler gap is about $\Delta a \sim 15\,\text{km} \simeq 3.2\, r_{\text{H,s}}$. The amplitudes of epicycle and vertical oscillations are comparable to $r_{\text{H,s}}$ and are about 1/3 and 1/2 of $\Delta a$, respectively. Therefore, the deviation of Daphnis's orbit from a circular and coplanar orbit is not negligible, and its effect on the satellite wakes and the vertical wall structures should be studied.

In fact, the Cassini observation revealed that the waveform of satellite wakes at the Keeler gap edge deviates from a sinusoidal waveform (Fig. 1(a); see also Porco et al., 2005; Seiß et al., 2010). Seiß et al. (2010) developed a collisionless streamline model considering an eccentric orbit of the embedded satellite and resonant perturbation by the satellite outside the ring system. They applied their model to the Daphnis-Keeler gap and the Pan-Encke gap systems. By fitting the observed wave patterns with their model, they estimated the Hill radius and non-zero eccentricity of the satellite. However, the obtained values deviate from the Cassini direct observation. One possible reason for the discrepancy is their neglect of particle-particle interactions.

Weiss et al. (2009) investigated the effect of the satellite eccentricity and inclination on the gap edge using a test particle model without any particle-particle interactions. They found that the satellite's orbital eccentricity and inclination would affect the gap edge wakes significantly and proposed that the out-of-plane perturbations of Daphnis create the vertical wall structure at the Keeler gap edge in the inclined orbit. On the other hand, our Paper I showed that the inelastic collisions between ring particles, with orbital eccentricities excited by the satellite perturbations, play the most crucial role in the formation of structures associated with the gap edge including the vertical wall. Therefore, in this paper, we investigate the effect of the satellite orbital eccentricity and inclination on the gap edge structures, taking account of particle-particle interactions by 3D global N-body simulation, as an extended study of Paper I.

Interestingly, it has recently been suggested that Daphnis's orbit could change with time in a relatively short timescale (Santana et al. (2019); Jacobson (2024); V. Lainey, 2023, private communication). The cause of this puzzling orbital change has not been understood yet. While Daphnis' eccentricity seems to be very small ($\ll r_{\text{H,s}}/a_s$) in an epoch of 2017, according to the Cassini final ring-grazing observation (Tiscareno et al., 2019, see also discussion in Section 3.3.1), the estimated eccentricity in an epoch of 2006 is relatively high ($\sim r_{\text{H,s}}/a_s$) (Jacobson et al., 2008; Seiß et al., 2010).

In Section 2, we briefly describe the simulation method and settings. Section 3 presents our main results with non-zero satellite eccentricity. We discuss its effect on the gap edge structures in detail and show that the satellite's orbital eccentricity is responsible for the detailed structures of the prominent mountain-range-like vertical walls casting shadows on the ring plane (Fig. 1). We also show that the satellite's orbital inclination raises non-local bending waves but does not form localized vertical wall structures. Section 4 consists of a summary of this paper.

## 2. Method and Simulation Settings

In the same way as Paper I, we use `n-body-with-center` based on Framework for Developing Particle Simulator (FDPS) (Iwasawa et al., 2016, 2020; Namekata et al., 2018) for our global N-body simulation. We use the soft-sphere model (e.g., Salo, 1995), where inelastic collisions are represented by adding a restoring harmonic force and an energy dissipation force during collisions (for a detailed description of the collision model, see Paper I). The restitution coefficient $\epsilon$ is set to be 0.1. The duration time of the collision (half of the restoring oscillation period) is set to be $T_{\rm col} = T_{\rm K}/(2^4 \cdot 2\pi) \sim 10^{-2} T_{\rm K}$ (see Appendix B), where $T_{\rm K}$ is the local Keplerian timescale at the Roche limit radius defined by $r_{\rm R} = 2.456 \left(\rho_{\rm p}/\rho\right)^{1/3} R_{\rm p}$, $\rho_{\rm p}$ and $\rho$ are bulk densities of the planet and the ring particles, and $R_{\rm p}$ is the physical radius of the planet (Saturn). The gravitational interactions between all the pairs of particles are calculated with the Barnes-Hut tree scheme (Barnes and Hut, 1986) available in FDPS. We integrate the equations of motion with the leapfrog method. The time step for the integration is $dt = T_{\rm K}/(2^{10} \cdot 2\pi) \sim 1.6 \times 10^{-4} T_{\rm K}$.

---

[1] In the case of the more massive satellite, Pan, the observed gap half width is $\Delta a \sim 8.8\, r_{\text{H,s}}$ (Table 1). The wider gap may be due to the long-term Lindblad torques, which makes the predicted epicycle amplitude excited by Pan at the gap edge smaller than that by Daphnis. As a result, the vertical wall should be less pronounced for Pan, which is also consistent with the non-detection of the vertical walls in the Pan-Encke gap system.

As in Paper I, $3 \times 10^6$ particles are set in global coordinates, in a radial range of $0.41\,r_{\rm R}$ to $0.82\,r_{\rm R}$ with a constant surface density, where $r_{\rm R}$ is the Roche limit radius. In our simulations, the bulk density and mass of the ring particles are given by $\rho = 0.5\,{\rm g/cm^3}$ and $m = 4 \times 10^{-11} M_{\rm p}$, respectively. The much larger individual mass of ring particles than that in the real system would not affect the results here because the vertical walls are formed by collective dynamics of ring particles excited by the satellite and the artificially enhanced collisional viscosity does not considerably exceed the effective viscosity due to the ring particle self-gravity wakes (Paper I). We assume that all particles are spherical and have an equal size determined by their mass and bulk density. The ratio of particle physical radius $r_{\rm p}$ and Hill radius of the satellite $r_{\rm H,s}$ is $r_{\rm p}/r_{\rm H,s} \sim 0.02$, and the ratio of Hill radius of the ring particle $r_{\rm H,p}$ and $2 r_{\rm p}$ is $r_{\rm H,p}/2 r_{\rm p} \sim 0.7$ in our simulation. The initial optical depth of the ring particles is $\tau \sim 0.1$.

The satellite mass $M_{\rm s}$ is set to be $5 \times 10^{-6} M_{\rm p}$, which is much larger than the actual Daphnis mass ($M_{\rm s} \simeq 1.2 \times 10^{-13} M_{\rm p}$). However, as already mentioned, the results for the dynamics driven by the scattering of the satellite are scaled by the satellite Hill radius. We will show that the simulation results re-scaled to a Daphnis-mass satellite match the Cassini direct observations.

In order to focus on the effect of the satellite eccentricity and inclination on the gap edge structure, we fix the satellite orbit to an eccentric and non-inclined orbit or an inclined and circular orbit with semimajor axis $a_{\rm s} = 0.66\,r_{\rm R}$. With $\rho = 0.5\,{\rm g/cm^3}$ and $\rho_{\rm p} = 0.69\,{\rm g/cm^3}$, $r_{\rm R} \simeq 2.73\,R_{\rm p}$. The assumed semimajor axis in the simulation is $a_{\rm s} = 0.66\,r_{\rm R} \simeq 1.80\,R_{\rm p}$. Because the actual semimajor axis of Daphnis is $2.26\,R_{\rm p}$, the assumed semimajor axis in the simulation is $\sim 20\%$ smaller.

The corresponding amplitude of radial excursion is $a_{\rm s} e_{\rm s} = 6.6 \times 10^{-3} r_{\rm R} \simeq 0.8 r_{\rm H,s} \simeq \Delta a/4.3$. According to the $r_{\rm H,s}$ scaling, eccentricity and inclination are given by half radial excursion and half vertical oscillation, respectively, scaled by $r_{\rm H,s}$ as

$$\tilde{e}_{\rm s} = \frac{e_{\rm s}}{(M_{\rm s}/3M_{\rm p})^{1/3}}$$

$$\simeq 0.91 \left(\frac{e_{\rm s}}{3.1 \times 10^{-5}}\right) \left(\frac{M_{\rm s}}{M_{\rm D}}\right)^{-1/3} \simeq 0.84 \left(\frac{e_{\rm s}}{0.01}\right) \left(\frac{M_{\rm s}}{5 \times 10^{-6} M_{\rm p}}\right)^{-1/3}, \quad (1)$$

$$\tilde{i}_{\rm s} = \frac{i_{\rm s}}{(M_{\rm s}/3M_{\rm p})^{1/3}}$$

$$\simeq 1.8 \left(\frac{i_{\rm s}}{6.3 \times 10^{-5}}\right) \left(\frac{M_{\rm s}}{M_{\rm D}}\right)^{-1/3} \simeq 0.42 \left(\frac{i_{\rm s}}{0.005}\right) \left(\frac{M_{\rm s}}{5 \times 10^{-6} M_{\rm p}}\right)^{-1/3}, \quad (2)$$

where the non-scaled inclination values are in radian, and $e_{\rm s} = 3.1 \times 10^{-5}$ and $i_{\rm s} = 6.3 \times 10^{-5}$ are the reference mean values of Daphnis (Table 1), and $M_{\rm D} = 1.2 \times 10^{-13} M_{\rm p}$ is the mean value of Daphnis mass (Weiss et al., 2009).

In the first run, an eccentric non-inclined satellite orbit with $(e_{\rm s}, i_{\rm s}) = (0.01, 0.0)$ is considered. The assigned eccentricity $e_{\rm s} = 0.01$ in the simulation with $M_{\rm s} = 5 \times 10^{-6} M_{\rm p}$ corresponds to $\tilde{e}_{\rm s} = 0.84$, which is comparable to the Daphnis' normalized eccentricity. In the second run, an inclined circular orbit with $(e_{\rm s}, i_{\rm s}) = (0.0, 0.005)$ is considered. We note that the assigned value of $i_{\rm s}$ corresponds to $\tilde{i}_{\rm s} = 0.42$, which is about 4 times smaller than $\tilde{i}_{\rm s}$ for the reference mean inclination of Daphnis. In our simulations, the vertical amplitude of bending waves induced by the satellite is much more exaggerated compared to the local vertical wall structure. Because our purpose here is to simulate the superposition of the global bending waves and the local vertical wall structure, and because the bending wave radial wavelength is independent of satellite inclination (Eq. (3)), we adopt 4 times smaller value of inclination than that by the $r_{\rm H,s}$ scaling for visibility.

## 3. Results
### 3.1. Effect of Satellite's Orbital Eccentricity
#### 3.1.1. Non-sinusoidal sawtooth-like satellite wakes

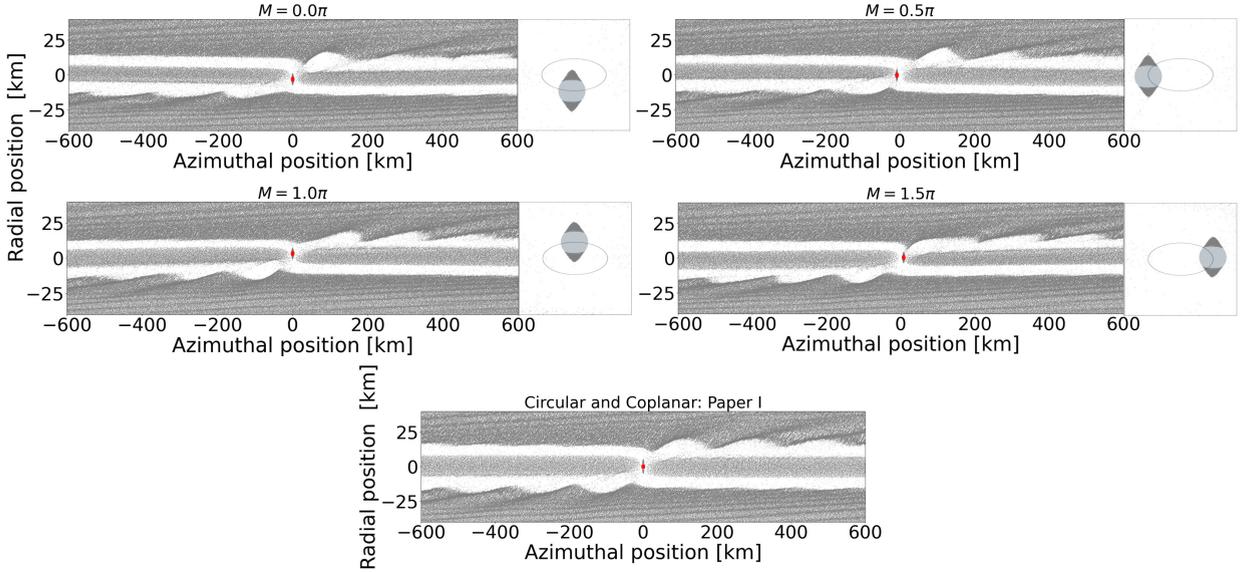

**Figure 2:** The snapshots of satellite wakes (left panels) and the zoom-in of the vicinity of the satellite with the epicycle depicted with a black line (right panels) at different times. $M = 0.0\pi$ corresponds to $t \simeq 127 T_{\rm K}$, where $t$ is the time in the simulation. For comparison, the snapshot with circular satellite orbit (Paper I) is shown in the bottom panel. The axes are converted to real scales at Daphnis orbit with the $r_{\rm H,s}$ scaling (see Section 1).

Here, we present a result of the simulation with the satellite in the fixed eccentric non-inclined orbit, $(e_{\rm s}, i_{\rm s}) = (0.01, 0.0)$. Figure 2 shows snapshots of satellite wakes at four different satellite epicycle phases in the radially local frame that is rotating with the epicycle guiding center of the satellite. We cut the ring region of the radial position in the range of $x \equiv r - a_{\rm s} = [-8.6, 8.6]\, r_{\rm H,s}$ and the azimuthal range of $\theta \simeq [-\pi/2, \pi/2]$ and stretched to the rectangular coordinates for visual convenience. The vertical and horizontal axes are radial and azimuthal positions relative to the satellite epicycle guiding center. In this figure, the scales of horizontal and vertical axes with $M_{\rm s}/M_{\rm p} = 5 \times 10^{-6}$ are converted to the real ones by the $r_{\rm H,s}$ scaling discussed in Section 1. The locations of the epicycle motion of the satellite in the rotating frame are equivalent to the mean anomaly of the satellite $M = n_{\rm s} t$ in the inertial frame, which is shown at the top of each panel, where $n_{\rm s}$ is satellite's mean motion and $t = 0$ is defined at the time when the satellite is at the planetocentric periapse. The satellite position, which is marked by a red dot in the left panels and zoomed in the right panels, rotates around the guiding center with the small radial amplitude of $\simeq 0.84\, r_{\rm H,s} \simeq 3.9\,{\rm km}$. For comparison, we also show the snapshot in the case of the circular satellite orbit (Paper I) in the bottom panel. We note that many particles remain in the horseshoe region, and these particles decrease with time (see Fig. 5 in Paper I). The detailed time evolution of horseshoe particles is out of the scope of this paper. We also note that many particles cluster around the satellite with non-zero eccentricity (right panels), which was also observed in Paper I. The evolution of this cluster and subsequent formation of a ridge near the equator of gap-opening satellite discovered by Cassini observation (see Buratti et al., 2019) will be investigated in a future work.

It is clearly shown that the waveforms created by the eccentric-orbit satellite are non-sinusoidal, in contrast to the case of the circular satellite orbit. The waveforms fluctuate periodically with time, and the fluctuation period is identical to the satellite's orbital period. The streamline models qualitatively predicted the time-varying non-sinusoidal waveform and also actually observed by Cassini (Fig. 1(a), (b); see also Figs. 1 and 9 of Seiß et al. (2010)). The observed sawtooth-like waveforms in the outer edge in Fig. 1(b) fit the waveforms in the outer edge at $M = 1.5\pi$ with the satellite in an eccentric orbit in our simulation shown in Fig. 2. The observed waveforms and scales in the upper and lower panels in Figure 9 of Seiß et al. (2010) perfectly fit the waveforms in our results at $M = 0.5\pi$ and $M = 1.5\pi$, respectively, in both edges. Both detailed wave patterns and spatial scales in the Cassini observation are reproduced.

Another effect of the satellite orbital eccentricity is shaping the vertical walls, as we discuss in Section 3.1.2. Before that, we briefly discuss the non-sinusoidal sawtooth-like satellite wakes. Our result is consistent with Seiß et al. (2010)'s results with the streamline model. Our simulation includes ring particles inelastic collisions and self-gravity, which were not included in the Seiß et al. (2010)'s streamline model. We found that both effects do not modify the sawtooth-like non-sinusoidal satellite wakes, while the inter-particle inelastic collisions sharpen the gap edge and the self-gravity excites the self-gravity wakes with much shorter wavelength than that of the satellite wakes (Section 3.3).

### 3.1.2. Azimuthally confined higher vertical walls

We investigate the effect of satellite eccentricity on the vertical structure of the satellite wakes. Figure 3 shows the snapshots of eccentricities and inclinations of ring particles in the satellite wake region at $x \simeq [0.0, 7.7] r_{H,s}$ at different epicycle phases of the satellite. For comparison, the snapshot in the case of a circular and coplanar satellite orbit is shown in the bottom panel. We find that the eccentricity distribution of the particles shows a time-varying damped oscillation pattern with $\theta$, meaning that the eccentricity excitation of the ring particles depends on the satellite epicycle phase (Cuzzi et al., 2024). The particle eccentricities are much more highly excited for the particles in the outer gap edge that pass the satellite near the planetocentric apoapse ($M \sim 1.0\pi$) and undergo closer encounters with the satellite, while the excitation is lower for the passing of the satellite near the planetocentric periapse ($M \sim 0.0$). Because the eccentricities of the particles are fully excited after $\sim T_K/4$, they are peaked around $M \sim 1.5\pi$. As we have proposed in Paper I, the ring particles' excited lateral motions (eccentricities) are converted to their vertical motions (inclinations) through particle-particle collisions. Accordingly, the distribution of the particle's inclination excitation also has peaks that synchronize with those in the eccentricity excitation (see also the movie in Movie S1), which indicates that a snapshot of the vertical wall structures is azimuthally more localized compared to that found in the case of the circular-orbit satellite (Paper I).

Figure 4 shows the optical depth normalized with initial optical depth $\tau_0$ at each grid on the $z$-$\theta$ plane. The bottom panel corresponds to the result of circular and coplanar satellite orbit stationary in the co-rotating coordinates. In the case of an eccentric-orbit satellite, particles are strongly splashed in localized regions (see also the middle panel of Fig. 3 a) around $M \sim 0.0 (= 2.0\pi)$ after $\sim T_K/2$ from the closest encounters at $M \sim 1.0\pi$. On the other hand, around $M \sim 1.0\pi$, the vertical structure is more similar to that in the circular and coplanar satellite orbit case. This indicates that the vertical walls in the outer gap edge are also not stationary and are higher and localized when the satellite is near the periapse ($M \sim 0.0$). This explains the observed azimuthally local feature that the shadows of the first and second wakes are clearly separated (see Fig. 1). We measured the vertical height with fitting the vertical profile of the optical depth with the sum of two Gaussian functions (see Appendix A in detail). The measured wall height is $\simeq 0.172 r_{H,s}$ when $M = 0.0\pi$ in the eccentric-orbit satellite case and $\simeq 0.145 r_{H,s}$ in the circular-orbit case. The higher wall in the eccentric-orbit satellite case ($\sim 0.2 r_{H,s}$) is also more consistent with the observation than the circular-orbit case ($\sim 0.1 r_{H,s}$). We note that the vertical height could depend on several parameters of the rings and the gap-opening satellite (e.g., ring optical depth, the restitution coefficient, the particle size distribution and the satellite eccentricity). We will investigate these dependencies on the gap edge structures in a future study.

### 3.2. Effect of Satellite's Orbital Inclination

Figure 5 shows the 3D plots of the satellite wakes created by the satellites with $(e_s, i_s) = (0.0, 0.005)$. The colors of particles indicate the $z$-component of their positions. The out-of-plane perturbation from the satellite in the inclined orbit induces a global fluttering of the ring plane (i.e. bending waves). On the other hand, the local vertical structures created through scattering by the satellite and particle-particle collisions persist at wavefronts even in this case with non-zero $i_s$.

Figure 6 shows the first, second, and third satellite wakes cross-sections. The left panel clearly shows that in the non-zero $i_s$ case, the local vertical structure by collisions and the global disk fluttering by the satellite's inclination is created simultaneously and superposed, preserving their intrinsic variation patterns. The height of the local vertical structure from the disk plane is $\sim 0.1 r_{H,s}$ in both the $i_s \neq 0$ case (the left panel) and the $i_s = 0$ case (the right panel). Thus, our global simulation indicates that particle-particle collisions at the wavefront create the azimuthally localized vertical wall structure but not by out-of-plane perturbation from the inclined-orbit satellite.

As shown below, if we apply Hahn (2007)'s argument for the Daphnis-Keeler gap system, the wavelength of the bending waves is $\sim 30$ times larger than $r_{H,s}$, which regulates the spatial scales of the structures of satellite wakes and the associated vertical walls, both in the real system and the simulation. These two structures coexist preserving their intrinsic variations.

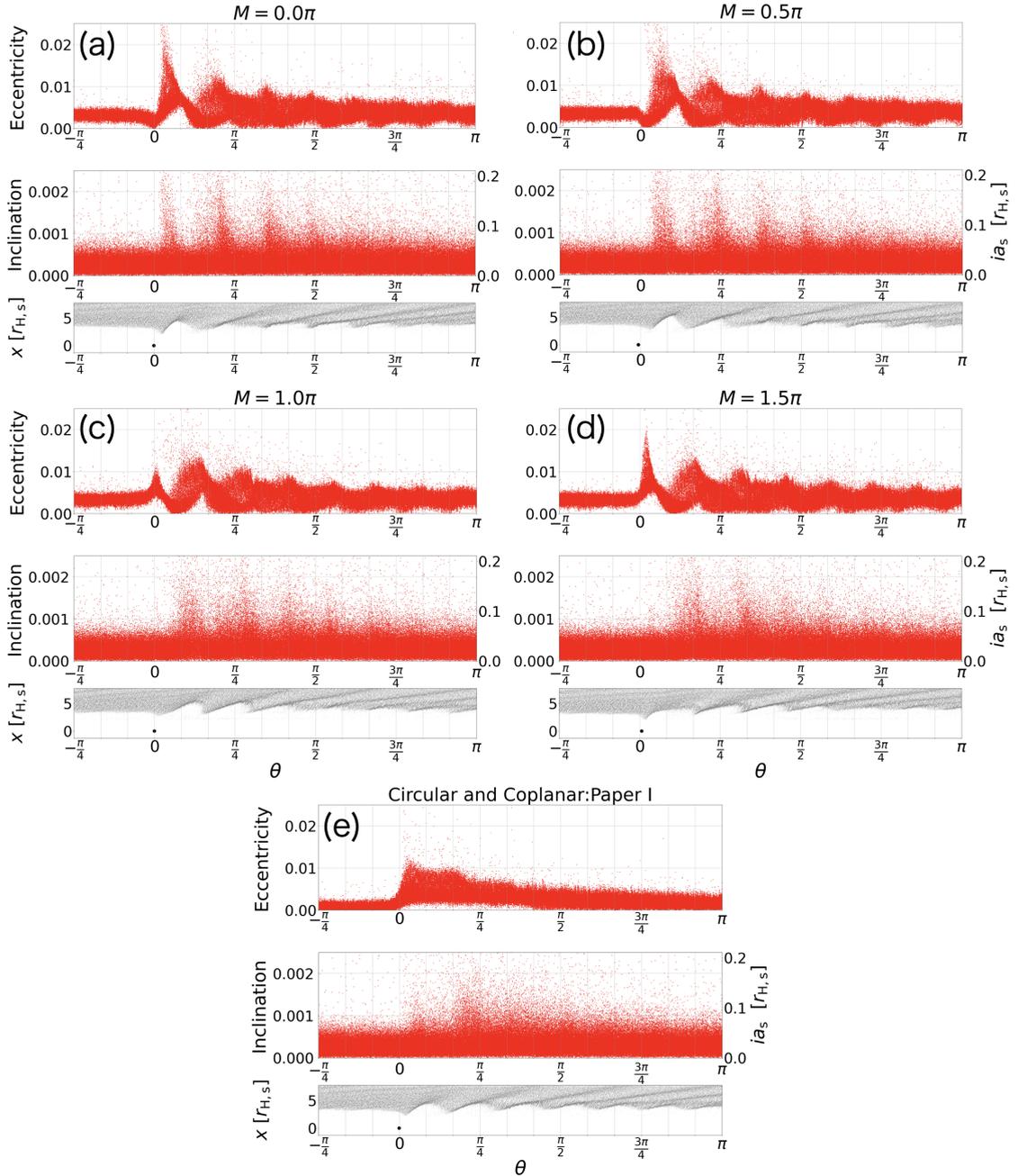

**Figure 3:** The snapshots of eccentricities (top panels of each plot) and inclinations (middle panels of each plot) of particles at the satellite wake region of $\tilde{x} \simeq [0.0, 7.7]$ at different times when the satellite mean anomaly is $0.0, \pi/2, \pi$ and $3\pi/2$. The bottom panels of each plot are the snapshot of the satellite wake region. For comparison, the plots in the case of the circular satellite orbit (Paper I) are also shown in panel (e). The movie version of this figure is shown in Movie S1 and S2.

Hahn (2007) investigated the secular perturbations from the satellite in an inclined orbit to a nearby gap edge and the excitation of the bending waves by modeling the disk as a nested set of gravitating rings (Hahn, 2003). The paper analytically derived its wavenumber $k$ at the disk exterior to the satellite orbit as a function of fractional distance $\alpha = |a - a_{\rm edge}|/a_{\rm s}$ in the limit in which the satellite mass is much smaller than the disk mass:

$$k \simeq \frac{1}{0.87\pi a\Delta}\left[2 + \frac{\mu_{\rm c}}{\mu_{\rm d}}\left(1 + \frac{\alpha}{\Delta}\right)\right], \qquad (3)$$

where $a$, $a_{\rm edge}$ and $a_{\rm s}$ are the radial distance from the central planet, the orbital radius of the gap edge and the orbital radius of the satellite, respectively, and $\Delta = |a_{\rm edge} - a_{\rm s}|/a_{\rm s}$ is the fractional distance between the satellite orbit and the disk's inner edge. The parameter $\mu_d$ is the disk mass normalized by the central planet's mass, and $\mu_c$ is defined as

$$\mu_c \equiv \frac{21\pi}{4} \left( \frac{R_p \Delta}{a_s} \right)^2 J_2, \tag{4}$$

where $J_2$ is the planet's second zonal harmonic. Equation (3) shows that the $\alpha$-dependence is weak and the bending waves are not radially confined. Using the total mass of the Saturn's ring measured by Cassini ring-grazing observation (Iess et al., 2019), $\mu_d \sim 2 \times 10^{-8}$, and the Daphnis mass, $M_s/M_p \sim 1.2 \times 10^{-13}$, we obtain $\mu_c/\mu_d \sim 0.02$ in the real system. With $\mu_c/\mu_d \sim 0.02$, the wavelength of bending waves at the gap edge ($\alpha = 0$) is $\lambda_{bend} \sim 30\,r_{H,s}$. The characteristic length scale of the satellite wakes and the associated vertical walls is $\sim \mathcal{O}(1)\,r_{H,s}$ and $\sim \mathcal{O}(0.1)\,r_{H,s}$, respectively.

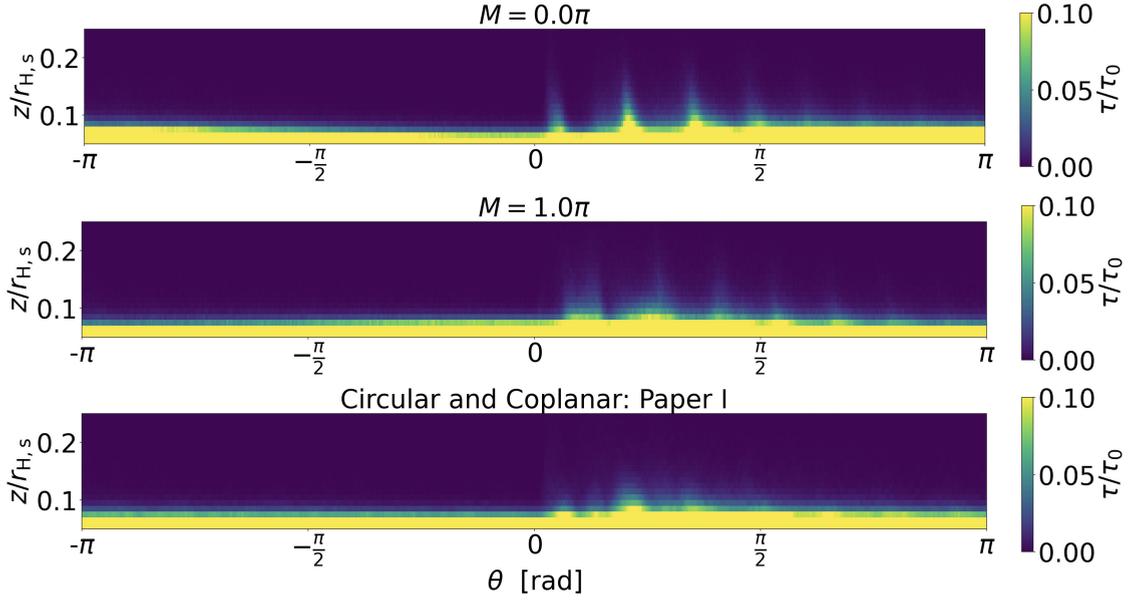

**Figure 4:** The $z-\theta$ plots of the optical depth normalized with initial optical depth $\tau_0$. The top and middle panels correspond to different times when the satellite's mean anomaly is $0.0$ and $1.0\pi$, respectively. The bottom panel shows the case of circular satellite orbit (Paper I).

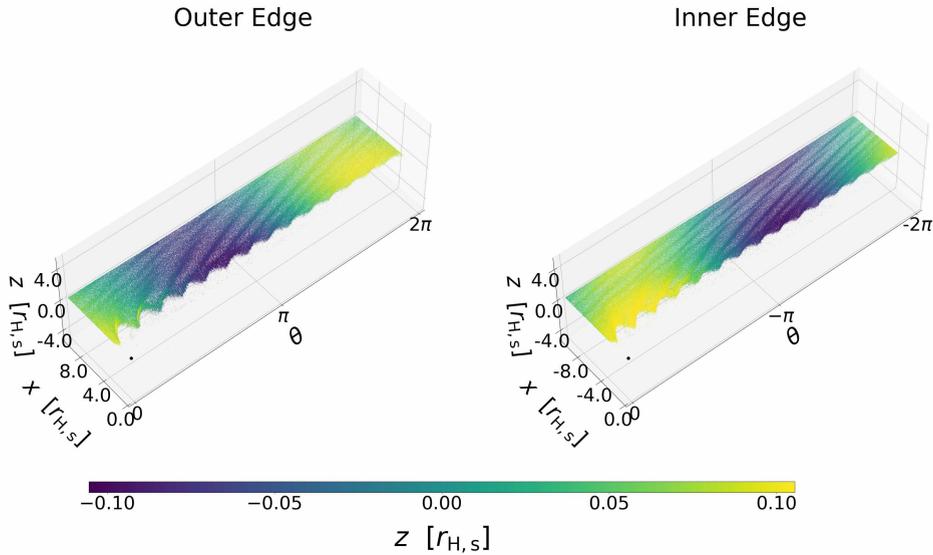

**Figure 5:** The 3D plot of the inner and outer satellite wakes in the case of the satellite inclination $5 \times 10^{-3}$ rad. Color shows the $z$-component of each particle.

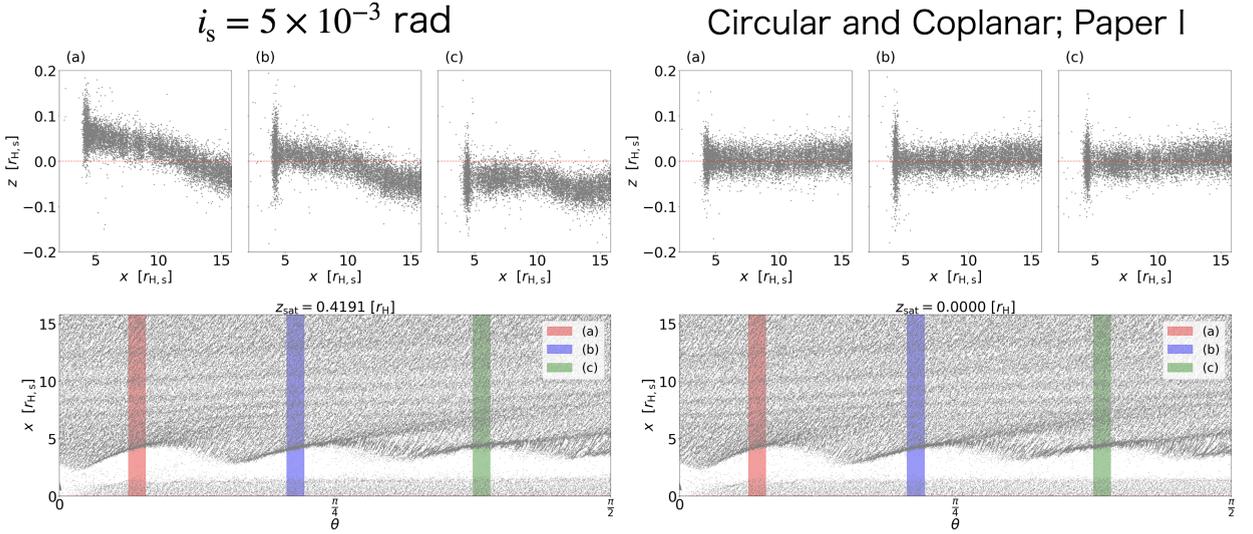

**Figure 6:** The cross-sections of satellite wake regions in the case of $i_s = 5 \times 10^{-3}$ rad (left panels) and circular and coplanar satellite orbit (right panels). (a), (b) and (c) correspond to the first, second, and third satellite wake regions shown as colored bands in the bottom plots.

This estimate would support our simulation result that the local vertical structures casting the observed shadows are created by the coupled effect of the satellite wakes and the inter-particle collisions in the wakes but not by the effect of the non-zero orbital inclination of Daphnis.

### 3.3. Notable features in comparison with the Cassini ring-grazing observation
#### 3.3.1. Orbit of Daphnis

Figure 7 shows a zoom-in view of detailed structures of satellite wakes by the Cassini ring-grazing observations and the corresponding simulation results with the non-zero eccentricity, as well as the simulation result with $e_s = 0$ (see Fig. 4 in Paper I). The observation image was taken in 2016–2017, which was later than the image in Fig. 1(a), (b) and Fig. 9 of Seiß et al. (2010).

The simulation result with the circular orbit of the satellite (panel b) best reproduces the observed satellite wakes (panel a) among the panels b to f, even for the detailed structures, as well as the structure scales. The interesting point is that the results (panels c to f) with the eccentric orbit of the satellite do not fit the observations at any epicycle phases. The bright white parts in the observed image (panel a) correspond to the vertical walls. The walls are more continuous than those in the eccentric orbit case of Fig. 1(b), which also supports the satellite orbit close to a circular one. Because the satellite wakes scale is $\sim \mathcal{O}(1)\,r_{H,s}$, the satellite's orbital eccentricity can be constrained from the waveforms in a resolution of $\sim r_{H,s}/a_s$. While $\tilde{e}_s \sim 1$ with the reference observation (Table 1), $\tilde{e}_s \ll 1$ is strongly suggested at the time of the Cassini ring-grazing observations according to the comparison of the observation image and our simulation results. As we already pointed out, $e_s$ of Daphnis could vary with time (Table 1, and references therein). We will focus on the details of this issue in a separate paper. A more detailed analysis of this possibility would be needed using the observation images and the simulation results.

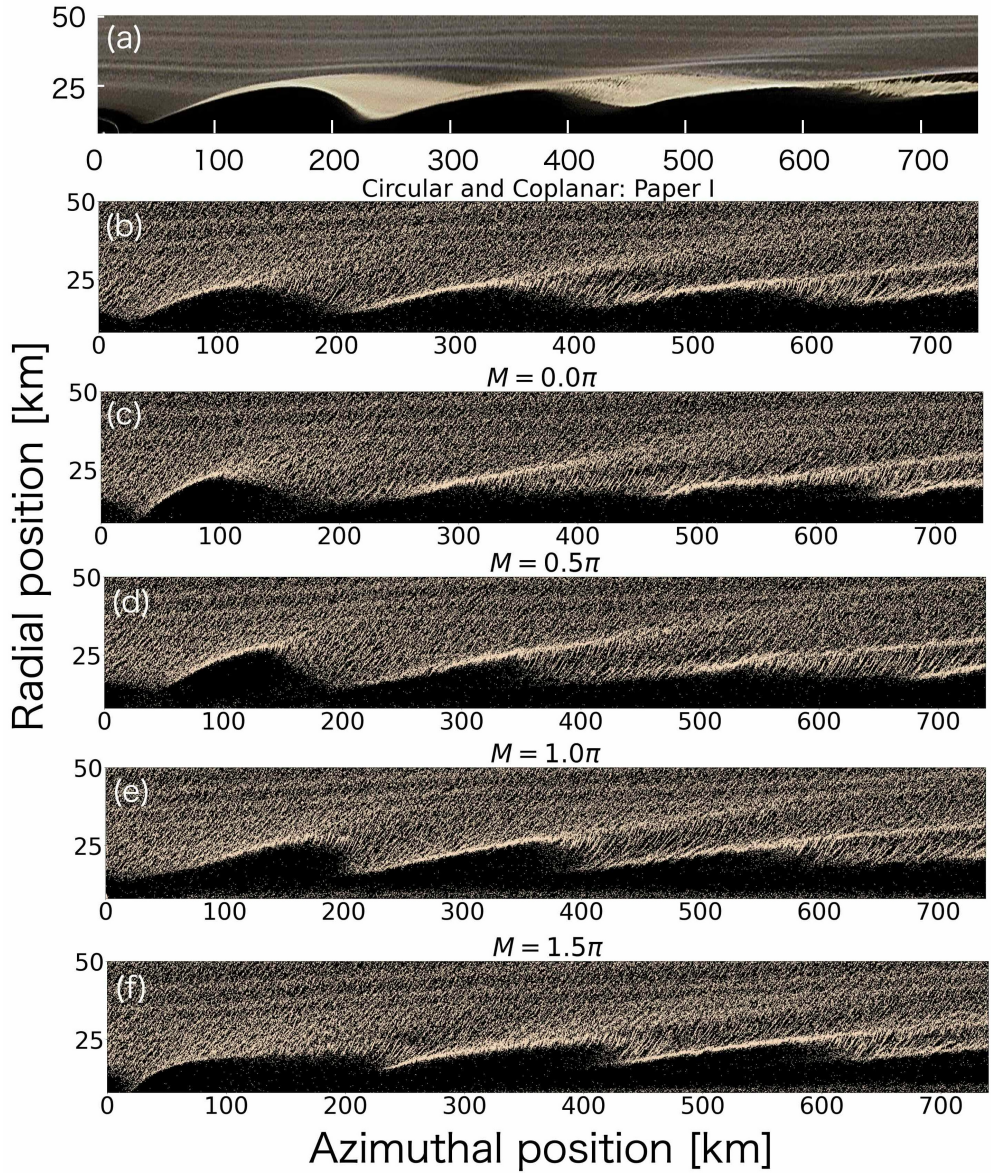

**Figure 7:** The comparison of the zoom-in satellite and self-gravity wake structures of the Cassini observation and N-body simulations. The images are compressed by a factor of 2 in the radial direction. Panel (a) is the image taken in Cassini's ring-grazing observation (https://www.jpl.nasa.gov/images/pia23167-embedded-moons-sculpt-saturns-rings/ ; CREDIT–NASA/JPL/Space Science Institute). Panels (c) to (f) show the zoom-in view of the satellite wake regions in Fig. 2 at $M = 0.0, 0.5\pi, 1.0\pi$ and $1.5\pi$, respectively. Panel (b) shows the satellite wake region's zoom-in view in the circular and coplanar satellite orbit. The panel (a) length scale is measured by Fig. 1A of Cassini observations in Tiscareno et al. (2019). The axes of the panels (b)-(f) are converted to real scales at Daphnis orbit with the $r_{\rm H,s}$ scaling (see Section 1). The movie version of this figure is shown in Movie S3 and S4.

### 3.3.2. Self-gravity wakes inside the satellite wakes

Another interesting feature is that self-gravity wakes have a much shorter wavelength than satellite wakes. The Cassini observations show structures like the self-gravity wakes, particularly in high-surface density regions in the satellite wakes. The wavelength of self-gravity wakes near the Roche limit is given by the Toomre critical wavelength

(Salo, 1995; Daisaka and Ida, 1999):

$$\lambda_{\rm cr} \sim \frac{4\pi^2 G\Sigma}{\Omega^2} = \frac{4\pi^2 \Sigma r^3}{M_{\rm p}}, \tag{5}$$

where $\Sigma$ is the surface density and $\Omega$ is the Kepler angular velocity. In the real system, the ratio of Toomre critical wavelength and ring particle radius is $\lambda_{\rm cr}/r_{\rm p} \sim 14$ for particle radius of $r_{\rm p} \sim 5$ m and $\lambda_{\rm cr} \sim 70$ m (see below), while it is $\lambda_{\rm cr}/r_{\rm p} \sim 6.6$ in our simulation, which agree with each other within a factor of $\sim 2$. Thus, the self-gravity wakes are resolved in our global simulation. The self-gravity wakes have longer wavelengths and are more pronounced in higher surface density regions ($\lambda_{\rm cr} \propto \Sigma$), such as gap edges. Figure 8 shows the detailed structures of the self-gravity wakes. Our simulations not only reproduce the self-gravity wakes in the satellite wakes but also predict that the self-gravity wakes develop out of the satellite wakes. The self-gravity wakes are folded at the wavefront due to the Kepler shear flow (see also Movie S5 and Movie S6). As a result, the surface density there becomes much higher than in the other regions, stronger gravitational instability occurs locally, and more prominent self-gravity wakes develop. Here, we roughly estimate the wavelength of self-gravity wakes both in and out of the satellite wakes in the real system, but the detail quantitative analysis is left for future work. Using the observationally inferred surface density of the A ring $\Sigma \sim 400$ kg/m$^2$ (Tiscareno et al., 2007), the wavelength of self-gravity wakes out of the satellite wakes is estimated as $\lambda_{\rm cr} \sim 70$ m. When we simply consider the surface density dependence of $\lambda_{\rm cr}$ given by Eq (5), the wavelength of the self-gravity wakes elongating from the wavefront is $\sim 3 \times \lambda_{\rm cr} \sim 210$ m considering that the surface density at the wavefront is about three times larger than that in the other region (see Fig. 15 of Paper I).[2] On the other hand, the spatial resolution of Fig. 7 (a) at the Cassini ring-grazing observations was 170 meters per pixel (Tiscareno et al., 2019), which is between the wavelength of self-gravity wakes in the satellite wakes ($\sim 210$ m) and that out of the satellite wakes ($\sim 70$ m). This indicates that the straw-like patterns observed in the satellite wake region (see Fig. 7 (a)) correspond to the longer-wavelength self-gravity wakes elongating from the high-surface density wavefront.

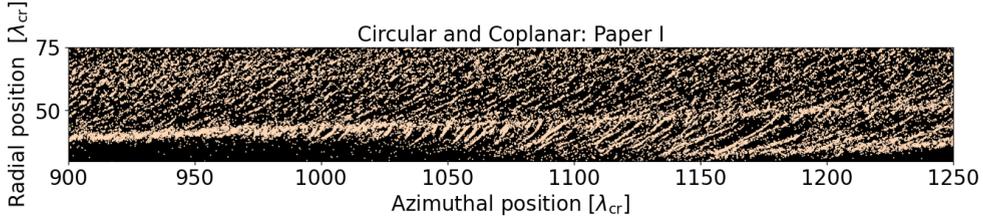

**Figure 8:** The zoom-in view of self-gravity wakes elongating from the wavefront of the third satellite wake and those out of the satellite wakes

in Figure 7 (b). The axis scales are normalized by $\lambda_{\rm cr}$ calculated with surface density $\Sigma$ in our simulation. The movie version of this figure is shown in Movie S5.

## 4. Summary and Discussion

The embedded small satellite, Daphnis, is opening a gap in Saturn's rings associated with several characteristic features both in the inner and outer gap edges, including satellite wakes, sharp edge, and spectacular vertical wall structures (e.g., Showalter, 1991; Porco et al., 2005; Weiss et al., 2009; Tiscareno et al., 2019). In our previous paper (Torii et al., 2024), we investigated the formation of these structures with a global 3D N-body simulation and reproduced them simultaneously, which have been treated with different theoretical approaches so far (e.g., Borderies et al., 1982, 1983, 1984, 1985, 1989; Lewis and Stewart, 2000, 2005, 2009; Lewis et al., 2011; Grätz et al., 2018, 2019; Weiss et al., 2009; Seiß et al., 2010).

In particular, Torii et al. (2024) revealed that the vertical wall structures are created by conversion of the particle's lateral motions excited by the satellite scattering into its vertical motions through particle-particle oblique collisions, which is in contrast to the previously proposed formation mechanism based on the out-of-plane perturbation from the satellite in an inclined orbit (Weiss et al., 2009). We predicted that the highest vertical wall is formed in the second wake and its height is $\sim 0.1\, r_{\rm H,s}$, which is consistent with the Cassini observation except that the prediction is a factor of 2 smaller than the observation.

---

[2]This may be underestimated because the shear should change in a strongly perturbed region such as in the satellite wake fronts (Stewart, 2022; Esposito et al., 2025).

However, our previous study only considered the case of circular and coplanar satellite orbit. In reality, Daphnis has the eccentricity and inclination such that their corresponding amplitudes of epicycle and vertical oscillations in a local frame rotating with the satellite guiding center are respectively as large as $\sim 1/3$ and $1/2$ of the gap width (Section 1). Thus, the satellite's epicycle (eccentricity $e_{\rm s}$) or vertical motion (inclination $i_{\rm s}$) should have important effects on the gap edge structures.

In this paper, as an extended study of Torii et al. (2024), we have focused on the effects of the $e_{\rm s}$ and $i_{\rm s}$ of the satellite orbit on the satellite wakes and its vertical structure. In the same way as in Torii et al. (2024), we have performed a high-resolution ($N = 3 \times 10^6$) 3D global N-body simulation of gap formation by the satellite in an eccentric or inclined orbit, considering all mutual interactions between ring particles including inelastic collisions and self-gravity. Although we used a much larger satellite and ring particles than the realistic ones for computational convenience, Torii et al. (2024) showed that the satellite wakes and its vertical structures can be scaled with satellite Hill radius $r_{\rm H,s}$. Hence, we extrapolate the simulation results to the actual systems. Accordingly, we adopted larger $e_{\rm s}$ (= 0.01) and $i_{\rm s}$ (= 0.005) estimated by the scaling argument from Daphnis ones. From the simulation with the eccentric-orbit satellite, we have found the followings:

1. Our simulation reproduces the time-varying non-sinusoidal satellite wakes due to the satellite's epicyclic motion (Fig. 2). The predicted non-sinusoidal features, as well as their spatial scales, match the image taken by Cassini (Fig. 1 a) and are qualitatively consistent with the previous streamline model (Seiß et al., 2010).
2. The vertical wall structure associated with the gap edge is also not stationary but periodically time-varying. The particle eccentricities in the outer gap edge are highly (weakly) excited when the satellite is near the apocenter (pericenter). Because the lateral motions of the particles induced by their eccentricities are converted into the vertical motions through particle-particle oblique collisions (Fig. 3), the wall height varies on the Keplerian timescale of the satellite.
3. As a result, azimuthally more localized and higher vertical wall structures are created by the eccentric-orbit satellite than by the circular-orbit satellite (Fig. 4), which is consistent with the fact that the observed shadows cast by the first and second satellite wakes are clearly separated (Fig. 1 b). The predicted wall height is $\sim 0.2\, r_{\rm H,s}$ in the case of $e_{\rm s} = 0.01$, which is higher by a factor of 2 than in the circular-orbit satellite case and more consistent with the observations.

From the simulation with the inclined-orbit satellite, we have found the followings:

1. The out-of-plane perturbation from the satellite induces a global fluttering of the ring plane (i.e., bending waves). In contrast, the local vertical structure due to the lateral scattering by the satellite persists at the wavefront of satellite wakes (Fig. 5 and Fig. 6).
2. The analytical formula Eq. (3) (Hahn, 2007) and our simulation show that in the real Daphnis-Keeler gap system, the length scale of the satellite wakes and its vertical structures are much smaller than the bending wavelength.

Thus, we have concluded that the vertical wall structure casting the shadows on the ring plane observed by Cassini is actually created by the particle's eccentricities due to the satellite's perturbations in an eccentric orbit and inter-particle collisions that convert the particle's lateral motions to the vertical motions, rather than the out-of-plane perturbation from the satellite in an inclined orbit. The vertical height of the wall possibly depends on several parameters such as ring optical depth, the restitution coefficient, the particle size distribution and satellite eccentricity and inclination. We will investigate these dependencies in a future study.

As in Section 3.3.1, the comparison of our simulations with the Cassini observations is consistent with Daphnis orbital eccentricity that varies with time. It was relatively large value ($\tilde{e}_{\rm s} \sim O(1)$) at least in an epoch of 2006 (Jacobson et al., 2008, see Table 1), while it may have been much smaller at the timing of Cassini ring-grazing observations in 2017. The cause of this puzzling orbital change of Daphnis has not been understood yet (Santana et al., 2019; Jacobson, 2024), thus it is worth studying further. It also suggests that the longer-wavelength self-gravity wakes ($\lambda_{\rm cr} \sim 210$ m) elongating from the satellite wavefront have already been detected by the ring-grazing observations. Still, the shorter-wavelength wakes out of the satellite wakes ($\lambda_{\rm cr} \sim 70$ m) were not detected due to the resolution limit at the ring-grazing observations ($\sim 140$ m per pixel). These may require more detailed analysis with N-body simulations and high-resolution observations. For the first point, it is easy to unlock the satellite orbit and investigate the satellite orbital variation with our global N-body simulation with considering any external perturbations such as, for example, resonant perturbation from the other satellite orbiting outside the A ring (e.g., Prometheus), which is left for a subsequent future work.

## 5. Acknowledgments

We thank Glen Stewart and Keiji Ohtsuki for their helpful and stimulating comments. Numerical computations were carried out on Cray XC50 at the Center for Computational Astrophysics, National Astronomical Observatory of Japan. This research is supported by JSPS Kakenhi grant 21H04512 and JST SPRING, Japan Grant Number JPMJSP2106.

## Appendix A. The measurement of the vertical height in the simulation results

In order to quantitatively compare the vertical wall shapes obtained in our simulations between circular and eccentric orbit cases, we define the vertical height of the wall structure. This also allows us to roughly compare our simulation results with observationally inferred the vertical wall height. We do not intend to propose a detailed method to estimate vertical wall height from observations of shadow length of the wall on the rings.

We calculate the optical depth normalized with the initial optical depth of the ring plane $\tau_0 \sim 0.1$ with averaging 165 snapshots at the same satellite epicycle phase. Then, we cut the $z - \theta$ plot at the azimuth corresponding to the first, second and third satellite wakes (shown as white dotted lines in the upper panels of Fig. 9 and Fig. 10) and obtain the vertical profile of the optical depth. Next, we fit the vertical profile of the optical depth with the sum of two Gaussian functions with different height (lower panels of Fig. 9 and Fig. 10):

$$f(z) = a_1 \exp\left\{-\frac{(z-z_0)^2}{2\sigma_1^2}\right\} + a_2 \exp\left\{-\frac{(z-z_0)^2}{2\sigma_2^2}\right\}, \tag{6}$$

where $a_1, a_2, z_0, \sigma_1$ and $\sigma_2$ are the fitting parameters and $a_1 > a_2$. The second term of Eq. (6) corresponds to the component of vertically splashing particles creating the wall structure. Finally, we define the vertical height as $3\sigma_2$. We note that this definition of the height is somewhat arbitrary. However, by introducing the definition, we can quantitatively compare the vertical height in our simulation results between circular-orbit and eccentric-orbit cases. It also allows us to roughly compare our simulation results to Cassini observation with this measurement. The measured values with this procedure are shown in the upper part of the lower panels.

## Appendix B. The dependence of the duration time of collision in the soft-sphere model

Here, we check the dependence of the duration time of collision $T_{\rm col}$ in our soft-sphere collision model on the formation of the gap edge structures. We set $T_{\rm col} = T_{\rm K}/(2^4 \cdot 2\pi) \sim 10^{-2} T_{\rm K}$ in all runs described above. We additionally performed three runs with smaller $T_{\rm col}$; $5.0 \times 10^{-3} T_{\rm K}$, $2.4 \times 10^{-3} T_{\rm K}$ and $1.2 \times 10^{-3} T_{\rm K}$ with the satellite fixed in a circular orbit. Fig. 11 shows the snapshots of the particle eccentricity and inclination as well as the snapshot of satellite wakes in each additional run. We find that the extent of the particle eccentricity excitation and vertical splashing is nearly the same in all runs. The particle's eccentricity is efficiently converted to its inclination, and the vertical wall forms regardless of the value of $T_{\rm col}$. Also, the dynamical features of the gap edge discussed here (e.g., gap width, wavelength of satellite wakes, self-gravity wakes, etc.) are hardly affected by the value of $T_{\rm col}$. Thus, we conclude that the duration time of collision is not a critical parameter for the formation of gap edge structures.

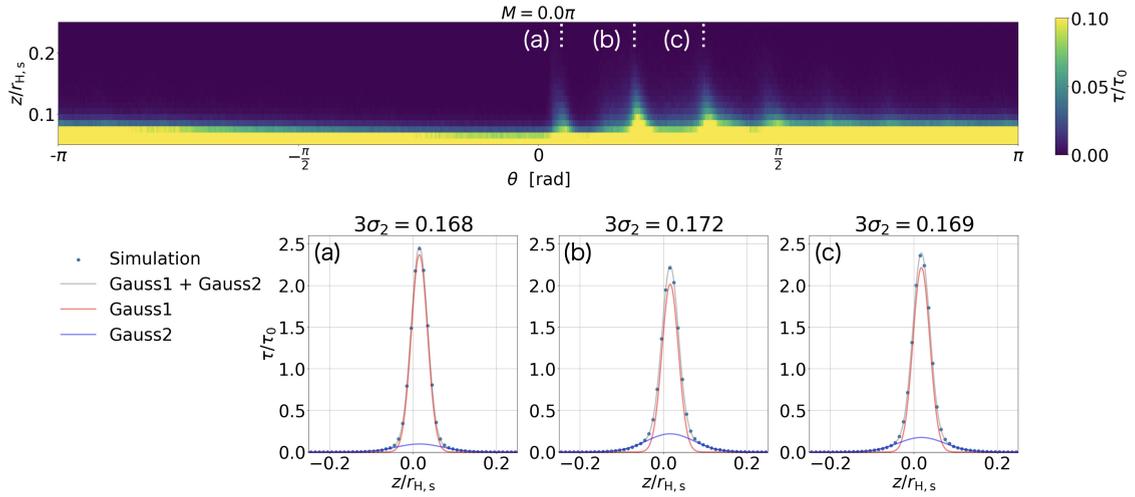

**Figure 9:** The upper panel shows the $z-\theta$ plot of the satellite wakes in the case of eccentric-orbit satellite case at $M = 0.0\pi$. The lower panels show the vertical profile of the optical depth (blue dots) and fitted Gaussian functions. The red and blue curves correspond to the first and second term of Eq. (6) and the grey curve corresponds to the sum of them. The values of measured vertical height defined as $3\sigma_2$ are shown in the upper part of the lower panels.

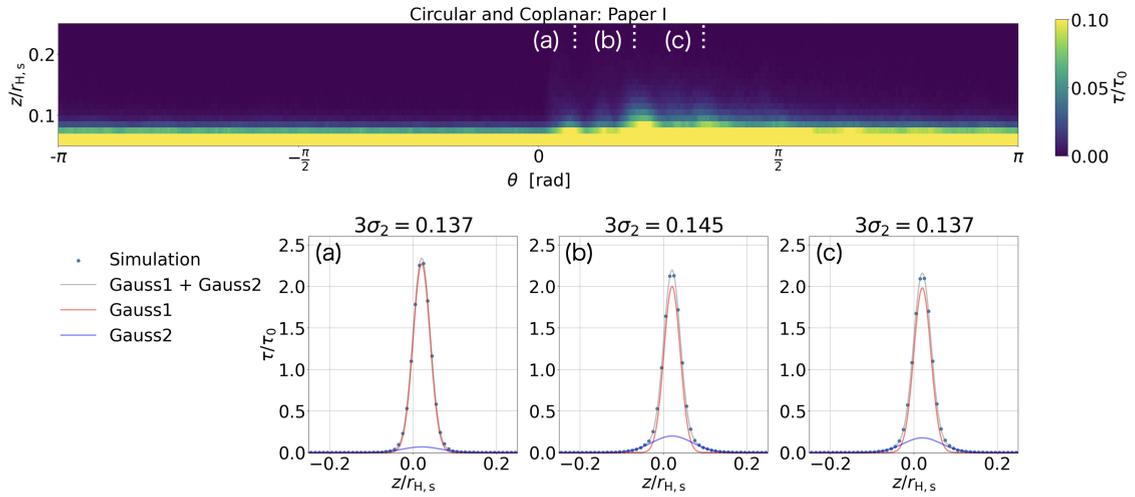

**Figure 10:** Same as Fig. 9 but for circular-orbit satellite case.

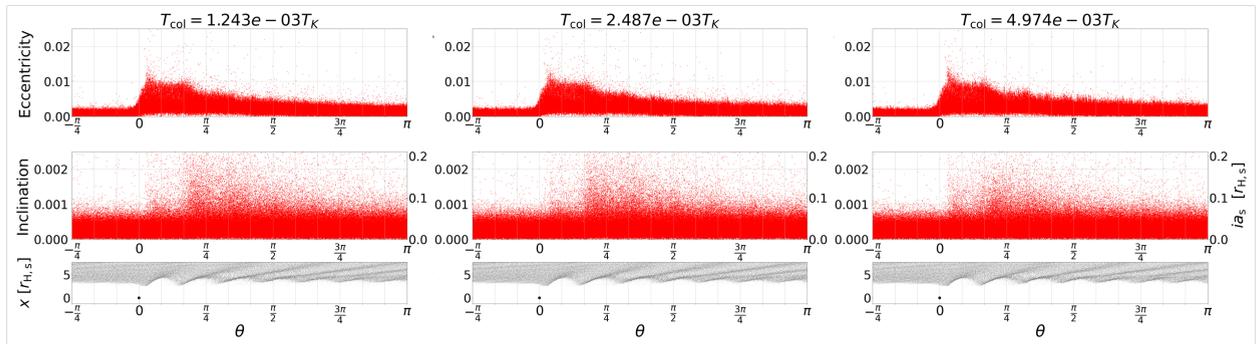

**Figure 11:** Comparison of additional runs with different duration time of collision $T_{\mathrm{col}}$.